\documentclass[12pt,preprint]{aastex}
\usepackage{graphicx}

\def\opeqn{\begin{equation}}
\def\cleqn{\end{equation}}
\def\lsim{\hbox{\rlap{\raise 0.425ex\hbox{$<$}}\lower 0.65ex\hbox{$\sim$}}}
\def\gsim{\hbox{\rlap{\raise 0.425ex\hbox{$>$}}\lower 0.65ex\hbox{$\sim$}}}

\begin{document}

\title{How Far Away are Gravitational Lens Caustics? \break  
                   Wrong Question}

\author{Sun Hong Rhie }



\begin{abstract}

It has been a persistent question at least for a decade where the gravitational 
lens caustics are in the radial direction: whether in front of the lensing mass, 
behind the lensing mass, or on the plane normal to the line of sight that passes 
through the lensing mass, the radiation source, or the observer. 
It is a wrong question. And, the truth angers certain referees who somehow possess 
the ability to write lengthy rubbish referee reports and delay certain papers 
indefinitely.   

General relativity is a metric theory, particularly of Riemannian geometry, 
which is characterized by the existence of an inner product -- or, the 
invariance of the proper time. According to Einstein field equations, a compact 
mass defines a spherical geometry around it and focuses photons from a distant 
source to an observer with the source and observer as the two focal points. 
When the mass is spherically symmetric, the two dimensional lens equation that 
relates the angular positions of a source and its images defines a point caustic 
at the angular position of the lensing mass.  
{\bf The third (radial) position of the point caustic is not defined.} For an 
arbitrary mass, the caustic extends into a web of piecewise smooth curves 
punctuated by cusps and again its notion exists only within the context of 
the lens equation. We point out a few errors in a couple of papers, published 
in the Astrophysical Journal, which may be influential.

\end{abstract}

\maketitle


\section{Spherical Geometry and Focal Points}
\label{secOne}

If we consider a 2-sphere, the geodesics (``straight lines") are the great
circles and any two great circles intersect in two places that are known as 
antipodal points in geography. There are no parallel geodesics on the sphere, 
and that is the well-known break-down of the fifth axiom (or postulate; 
see section\,\ref{secAxioms}) of plane geometry (or Euclidean geometry).
Given a pair of antipodal points such as north pole and south pole, there are 
infinitely many geodesics that diverge from one antipodal point and converge 
to the other.  The antipodal points are the foci of the geodesics. 

The 2-sphere ($S^2$) is characterized by a constant Gaussian curvature, 
and Gaussian curvature is nothing but the Riemann curvature scalar where
the non-unity  constant between them is just a matter 
of historical conventions. In two dimensions, Riemann curvature tensor
has only one independent component, and the Einstein tensor is given 
in terms of the Riemann curvature scalar and ``trivial tensor structure"
built from metric components. For the homogeneous space $S^2$,
the scalar curvature is constant -- Gauss curvature.   
(Weinberg\footnote{{\it Gravitation and Cosmology: Principles and Applications
of the General Theory of Relativity}, Steven Weinberg, John Wiley \& Sons, Inc.,
1972; Weinberg henceforth.}, p.144; Weinberg hereafter)

Just for a moment, let's stipulate for the purpose of illustration 
that the longitudes on $S^2$ are photon paths between a radiation 
source at the north pole and an observer at the south pole. 
A  image of an object an observer sees in its detector 
(say an array of light receptors including counters with $100\%$ 
efficiency, perfect homogeneity, and infinite resolution) is a 
distribution of photons on the detector that last-scattered the 
object (the last-scattering surface either of a radiation emission source 
that generates the photons, or of a reflecting object such as a planet,
etc) whose image the observer seeks. Thus, if we assume that the radiation
emission source at the north pole emits photons isotropically, the image 
of the radiation emission source (of the radius of a ``point")
that the observer at the south pole sees in its detector will be 
a circle (unless the detector is set exactly at the focal point and is 
burned; if the emission is directional such as in lasing, the image
will not be a circle because most of the geodesics that connect the
source and observer will not be followed by the photons from the 
``point" source.)    

 When the observer visits the north pole, the observer will find 
through meticulous measurements around the north pole 
that the radiation source has the size of a ``point" and the shape
of the radiating surface is not a circle. If we allow the observer
and source for the third dimension around the north pole for the sake
of imagination, the observer will be able to take a mugshot of 
the radiation surface which will show up as a ``point" in its
detector. In fact, the observer will most likely bring a lab-prepared
isotropically emitting ``point" light source to set up at the north 
pole and confirm that the ``point source" indeed produces a circular 
image when measured at the south pole.   

This dimensional change from a zero-dimensional point source to a 
one-dimensional circular image results in the point caustic of 
a spherically symmetric lensing mass in the context of the lens 
equation as we shall review in the following section.  

It should be useful to note that focal points are also referred to as 
conjugate points. For example, two antipodal points are a pair of conjugate
points; Or, the south pole is the conjugate point of the north pole.

\subsection{Axioms of Euclid Geometry}
\label{secAxioms}

We list the five axioms of Euclid which can be found in Euclid's textbook 
{\it Elements} written around 300 BC and had been used well into the 20th       
century.
\begin{enumerate}
   \item
     A straight line can be drawn from any point to any point.
   \item
     In a straight line, a finite straight line can be produced continuously.
   \item 
    A circle with any center and radius can be described by introducing
    a straight line segment as the radius and one of its end points as
    the center. 
   \item
    All right angles equal one another.
   \item
    If a straight line falling on two straight lines makes the interior 
angles on the same side less than two right angles, the two straight lines, 
if produced indefinitely, meet on that side on which are the angles less 
than the two right angles. 
\end{enumerate}

In order to avoid causing unnecessarily rigid authoritative impression
of the five postulates (or axioms) of Euclid, we remark 
that the axioms of plane geometry have evolved.
For example, seek the axioms by George Birkhoff who is perhaps better 
known for Birkhoff theorem (that the metric of a
spherically symmetric mass is static and given by Schwarzschild
metric). We should further note that Euclid was reluctant to use the
fifth axiom and his postulates are sometimes known as 4+1 postulates.
Considering the length of the fifth axiom, it is tempting to jump to 
a ``fuzzy axiom" that {\it simplicity is the beauty of truth}.

\section{The Lens Equation of a Spherically Symmetric Mass}

\subsection { Newtonian Deflection Angle }

In 1911, following up on his 1907 article on the gravitational 
influence on the propagation of light,\footnote{ 
A. Einstein, Jahrbuch f\"ur Radioact. und Elektronik, 4, 1907.}
Einstein published a calculation of the gravitational deflection 
angle by the Sun in an article titled 
``{\it \"Uber den Einfluss der Schwerkraft 
auf die Ausbreitung des Lichtes}," {\it Annalen der Physik}, 35.
\footnote{ 
``On the influence of gravitation on the propagation of light" (translation) 
in ``The Principle of Relativity: A collection of  original memoirs on the 
special and general theory of relativity," H.A. Lorentz, A. Einstein. 
H. Minkowski, and H. Weyl, Methuen \& Co, LTD., 1923.}  
Einstein got the deflection angle too small by factor 2 by considering the 
gravitational effect on the clock correctly but omitting the effect on the 
measuring rod. In other words, he got Newtonian deflection angle. It was 
before the systematic exposition of the general relativity culminating in 
what are now referred to as Einstein field equations (1916). We take it as a
historical lesson that we resort to the Einstein field equations whenever in 
doubt.

The masslessness of the light particles or $U(1)$ gauge invariance was not 
known in the era of Newton, but Newton's equation of motion of two particles 
under the mutual gravitation in the center of mass coordinates can be 
``naturally" extended to describe the motion of a massless photon under 
the gravitational influence of a mass $M$. 
If the masses of the two particles are $m_1$ and $m_2$, and their
reduced mass is $m: \ m^{-1} = m_1^{-1} + m_2^{-1}$, the motion of
the reduced mass $m$ (with relative position vector $x = r \hat r$)
is given by two equations in spherical coordinates 
where the scattering plane is chosen by $\theta = \pi/2$.
\opeqn
  m\,E = {m\over 2} \dot r^2 + {m\over 2} {J^2\over r^2} - {m\,GM\over r} 
    \ ; \qquad
  m\,J = m\,r^2 \dot\varphi  \ , 
\label{eqNewton}
\cleqn
where $E$ and $J$ are the (conserved) energy and angular momentum per unit 
reduced mass, and $M$ is the total mass. 

For the motion of photons, divide the equations with the reduced mass 
$m \neq 0$ and consider the resulting equations in the limit of 
$m \rightarrow 0$.  If impact parameter is denoted $b$ (as usual), 
\opeqn
  b \equiv \lim_{t\rightarrow -\infty} 
       r \sin\, (\varphi - \varphi_-)
\cleqn
where $\varphi_- \equiv \varphi(t=-\infty)$. The energy and angular
momentum of the photon can be determined from $b$ 
and eq.\,(\ref{eqNewton}) with $t=-\infty$.
\opeqn
  E = {1\over 2} \ ; \qquad J = b \ .  
\cleqn
The stationary condition $\dot r =0$ has one solution because $E >0$, and that 
is the distance ($r_0$) of the closest approach of the photon to the mass $M$. 
\opeqn
 r_0 = -GM + (G^2 M^2 + J^2)^{1/2} 
\cleqn
Since $J > 0$, the azimuthal angle increases with time ($\dot \varphi  > 0$). 
 
The equation of the orbit can be found by a simple integration of 
eq.\,(\ref{eqNewton}) after eliminating $dt$.  
For the incoming photon, 
\opeqn
 \int_{\varphi_-}^\varphi d\varphi
   = \int_\infty^r - {dr\over r^2}
      \left(1 + {2GM\over r} - {J^2\over r^2} \right)^{-1/2} \ ,
\cleqn
and the equation of the orbit is given by
\opeqn
 \varphi - \varphi_- = \cos^{-1}\left({-B\over A}\right)
        -\cos^{-1}\left({r^{-1}-B\over A}\right)  \ , 
\cleqn
where $B \equiv GMJ^{-2}$  and $A \equiv (J^{-2} + B^2)^{1/2}$.
If $\varphi_0 \equiv \varphi(r_0)$,  the photon trajectory is 
reflection symmetric with respect to the periastron, and its 
deflection angle is given by 
\opeqn
 \delta \varphi = 2(\varphi_0 -\varphi_-) - \pi
   = 2 \cos^{-1}\left({-B\over A}\right) - \pi \ .
\cleqn
The equation of the orbit can be rewritten in terms of $\varphi_0$
and results in the standard equation of a conic.
\footnote{ {\it Classical Mechanics},
H. Goldstein, 1965, Chap. 3; Goldstein hereafter.}
\opeqn
 {1\over r} = {GM\over J^2}\left( 1 + e \cos(\varphi-\varphi_0)
                           \right) \ ; \qquad
        e \equiv \left(1+ { J^2\over G^2 M^2} \right)^{1/2}
\cleqn 
where $e$ is the eccentricity of the conic section.
$e > 1 \ (E =1/2 > 0)$ and the orbit is hyperbolic.
In terms of the eccentricity,
\opeqn
 \delta\varphi = 2 \cos^{-1}\left(-{1\over e}\right) - \pi \ , 
\cleqn 
and for $J << GM$, in the linear order of $GM/J $,  
\opeqn
  \delta\varphi = {2 GM\over J} = {2 GM\over b} = {2 GM \over r_0} \ .   
\cleqn
This is the Newtonian deflection angle of the photon trajectory
due to  mass $M$. It is inversely proportional to the distance 
of the periastron $r_0$ correctly, but the numerical factor is wrong
by factor 2. The correct formula is $\delta\varphi = 4GM/r_0$
and can be obtained from Einstein field equations.

({\it The scope of the notations $A$ and $B$ used in this subsection
solely for the purpose of visual compactness of the expressions is
limited to this subsection.})

\subsection {Einstein Deflection Angle}

The Schwarzschild metric was found in 1916 promptly after the publication 
of Einstein's general relativity (Weinberg, Chap. 8), and Birkhoff theorem 
states that the metric of any spherically symmetric mass is static and given 
by the Schwarzschild metric (Weinberg, Chap. 11). Thus, we prefer to derive 
the Einstein deflection angle by a spherically symmetric mass $M$ by using
the exact metric solution of a mass $M$ instead of the Einstein's method of 
approximation in the 1916 thesis. The standard form of the Schwarzschild 
metric is given by
\opeqn
   ds^2 = -B dt^2 + A dr^2 + r^2 d\theta^2 
                  + r^2 \sin^2\theta d\varphi^2 \ ; \qquad
          B = A^{-1} = 1 - {2GM\over r} \ .  
\cleqn
A photon trajectory is a null geodesic in the metric, and the equation of 
motion (geodesic equation) can be obtained by the variational principle from 
the path integral of the Lagrangian ${\cal L}$. 
\opeqn
  S = \int {\cal L} dp = \int {1\over 2} g_{\mu\nu} \dot x^\mu \dot x^\nu dp \ ,
\label{eqAction}
\cleqn
where $g_{\mu\nu}$ is the metric tensor component, 
$\dot x^\mu \equiv d x^\mu/dp$, and $p$ parameterizes the path $x(p)$.
We can define the canonical momenta
\opeqn
  P_\mu \equiv {\partial {\cal L}\over \partial \dot x^\mu} \ , 
\cleqn
and N\"other's theorem guarantees that the time component $P_t$ and the 
azimuthal component $P_\varphi$ are conserved along the motion since the 
Lagrangian is independent of $t$ and $\varphi$. We also can define Hamiltonian
and it is conserved because the Lagrangian does not depend on the path 
parameter $p$ explicitly.
\opeqn
  {\cal H} \equiv P_\mu \dot x^\mu - {\cal L} = {\cal L}
   = {1\over 2} P_\mu P^\mu = {1\over 2} {ds^2\over dp^2} \ , 
\cleqn
The last equality shows that the invariance of the Hamiltonian is nothing but 
the invariance of the proper time $d\tau$. For the massless photons, 
$0 = d\tau^2 = - ds^2$, and $P^2 = P_\mu P^\mu = 0$.
\begin{eqnarray}
 P_t & = & - B \, \dot t = {\rm constant} \equiv - E
\label{eqPt} \\
 P_\varphi & = & r^2 \sin^2\theta \, \dot\varphi = {\rm constant} \equiv E J \\
 P_\theta & = & r^2 \dot\theta \\
 P_r & = &  B^{-1} \dot r \\
 P^2 & = & -B \dot t^2 + B^{-1} \dot r^2 + r^2 \dot\theta^2
       + r^2 \sin^2\theta \, \dot\varphi^2 = {\rm constant} = 0
\label{eqMomenta}
\end{eqnarray}
Since the metric is isotropic, choose an orbital plane by setting 
$\theta = \pi/2$ \ ($\dot\theta = 0$). 
Eliminate $dt$ and $dp$ from the equations of $P_t$, $P_\varphi$, and $P^2$, 
and integrate the resulting equation to obtain the equation of the orbit. 
The procedure is similar to that of the Newtonian scattering in the previous 
subsection (except that the integral can not be expressed in terms of elementary
functions) and can be found in any textbooks on general relativity 
({e.g.,} Weinberg, Chap. 8).  
In the linear regime $r >> GM$, we obtain the Einstein deflection angle
\opeqn
  \delta\varphi = {4GM\over r_0} \ , 
\cleqn
where $r_0$ is the distance of the periastron of the photon trajectory. 
It is worth noting that $g_{tt} = -B$ and $g_{rr} = A^{-1}$ equally contribute 
to the Einstein deflection angle, and the Newtonian deflection angle amounts 
to the contribution from $g_{tt}$. The factor 2 discrepancy between the 
Newtonian and Einstein deflection angles are the well-known general relativistic
factor 2, which was tested in 1919 during the eclipse notably by Eddington 
and his crew and numerously since.  

It should be useful to note that the variation of the action $S$ in 
eq.\,(\ref{eqAction}) leads to the ``standard" geodesic equation,
\opeqn
 0 = \ddot x^\lambda + \Gamma^\lambda_{\mu\nu} \dot x^\mu \dot x^\nu \ ,
\cleqn
where $\Gamma^\lambda_{\mu\nu}$ is the affine connection. Hence the path 
parameter $p$ is a so-called affine parameter which is a linear function of the
proper time. If we choose an arbitrary parameter, the equation develops extra
terms as one can check easily.

\subsection {The Lens Equation}

The source star and the observer are far away from the lensing mass $M$, and 
there the metric is effectively flat. The coordinates have been chosen such 
that $B, \, A \rightarrow 1$ as $r\rightarrow \infty$, hence the observer 
should feel relaxed to use the familiar flat space coordinate systems to make 
local measurements or to chart the sky knowing that the coordinate systems 
are valid all the way from the observer's neighborhood to the neighborhood of 
the source star except inside the star. 
In the asymptotically flat coordinate system, the photon 
arriving at the observer's detector after a long flight along a null geodesic 
would seem to come from a position in the sky that differs from the position 
of the source star where the latter is determined by hypothetically turning 
off the gravity by setting the Newton's constant $G=0$. The relation between 
a source position and its images in the observer's sky is the lens equation.

\begin{figure}
 \plotone{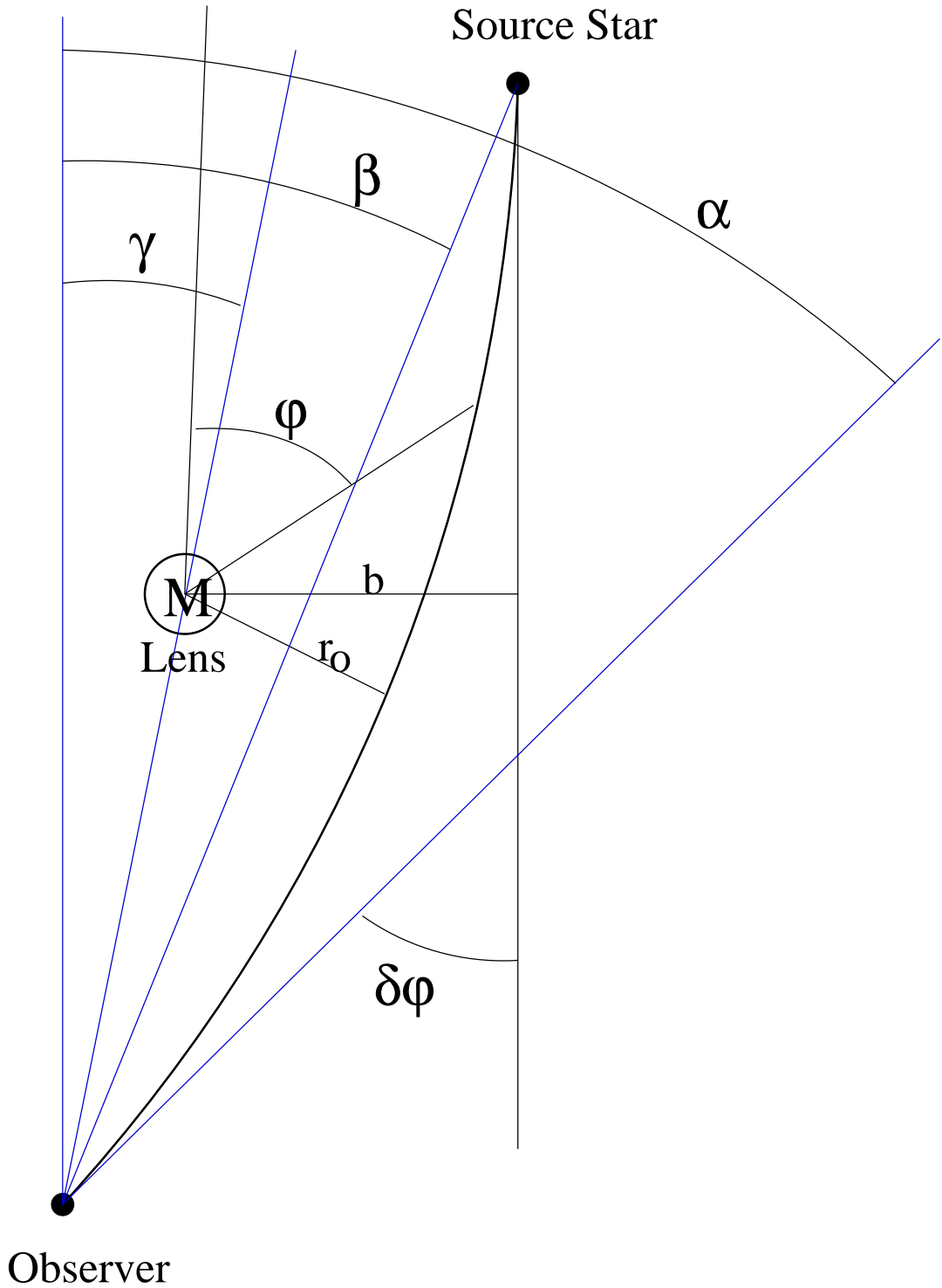} 
 \caption{Overlay of an observer's sky on the orbital plane $\theta=\pi/2$:
  $b$ is the impact distance, $r_0$ is the periastron distance, 
and $\delta\varphi$ is the deflection angle of the photon 
trajectory. If the source star is at the angular position $\beta$,
the observer sees an image corresponding to the depicted photon
trajectory at the angular position $\alpha$. The lens equation 
shows that there are either two images or two half-circle images,
forming a full circle, for a given source position $\beta$. 
The radius of the circular image is known as Einstein ring radius.} 
\label{fig_lens}
\end{figure}

Fig.\,\ref{fig_lens} shows an overlay of the observer's sky on the orbital plane
$\theta = \pi/2$ and a photon trajectory (null geodesic) that connects the 
source and the observer.  The position angles of the lens, source star, and 
image are depicted as $\gamma$, $\beta$, and $\alpha$, and they are related
to $r_0$ and $\delta\varphi$ by the followings.
\begin{eqnarray}
 & b  =  D_\ell \sin (\alpha-\gamma) \ ;
\label{eqLensOne} \\
 & D_s \sin (\alpha - \beta) = \tan \delta\varphi
    \left(  D_s \cos (\alpha - \beta)
           - D_\ell \cos (\alpha - \gamma)
           - b \tan (\delta\varphi / 2) \right) \ ,
\label{eqLensTwo}
\end{eqnarray}
where $D_\ell$ and $D_s$ are the (radial) distances to the lens mass $M$ and the
source star from the observer.  In the linear order small angle approximation,
the equations (\ref{eqLensOne}) and (\ref{eqLensTwo}) become
\opeqn
  b = D_\ell (\alpha-\gamma) \ ; \qquad
  D_s (\alpha - \beta) = \delta\varphi (D_s - D_\ell) \ .
\cleqn
Using the Einstein deflection angle $\delta\varphi = 4GM/r_0$
and $r_0 = b - GM \approx b$, 
the familiar  single lens equation is obtained.
\opeqn
  \alpha - \beta = {4GM\over \alpha - \gamma}
           \left({1\over D_\ell} - {1\over D_s} \right)
           \equiv {\alpha_E^2 \over \alpha - \gamma}
\label{eqLens}
\cleqn
Given the angular positions of a source ($\beta$) and a lens ($\gamma$), there 
are usually two solutions for $\alpha$;  the source, lens, and two images are 
collinear in the sky.  If the periastron distance $r_0 = R_E=D_1 \alpha_E$, 
however, the lens and source are aligned along the line of sight 
($\beta = \gamma$), and the image forms on a ring due to
the symmetry around the axis connecting the observer and the mass $M$.
The angular radius of the ring image is given by $\alpha_E$.
(There are two solutions $\alpha = \pm \alpha_E$ for $\theta = \pi/2$,
and the axial symmetry allows two solutions for each value of
$\theta = [0, \pi]$ resulting in solutions of two half circles
of the same radius $\alpha_E$. See section\,\ref{secBreak} 
for a way to understand
the transition from two images to two half-circle images.) The circular image 
is known as Einstein ring, hence the subscript $E$ in $\alpha_E$. 

The gravity of the mass $M$ makes the geometry of the space around it spherical
and focuses photons from the source to the observer; The source and observer 
are the two focal points of the geodesics that connect them. There are usually 
two geodesics that connect the source and observer, hence two images of a given 
source.  When the source and lens are aligned (as seen from
the observer with $G=0$), infinitely many null geodesics connect the source
and the observer, similarly to the longitudes connecting the south and north
poles on $S^2$ discussed in section\,\ref{secOne}; 
The source and observer are the two focal points of the infinitely many 
photon trajectories, and the observer sees a circular image of the source.  

In a following section, 
we shall see that the Einstein ring is also the critical curve. 
The caustic is by definition the curve  onto which the critical curve is mapped,
and the caustic of the single lens is a point caustic since the entire critical
curve is mapped to one point under the lens equation.
 In fact, the point caustic is a degenerate cusp. A cusp is by definition
 a point onto which a precusp -- a critical point where the critical eigenvector
 is tangent to the critical curve -- is mapped.

\subsubsection{A Diversion: Einstein Ring or Chwolson Ring?}

We find in monograph ``Gravitational Lenses", P. Schneider, J. Ehlers, and 
E. E. Falco, 1993, p.\,4 that Chwolson {\it remarked} in  ``{\it \"Uber eine
m\"ogliche Form fiktiver Doppelsterne}" of a ring-shaped image of the 
background star centered on the foreground star. Schneider et al. states that 
the circular image should be called Chwolson ring instead of Einstein ring.

Chwolson's main concern in the short article was a spectroscopic double star 
made of the foreground star and one (faint) image of the background star near 
the foreground lensing star. We recognize that the title ``On a possible form
of fictitious double stars" may invoke in the minds of today's readers of a 
spectroscopic double star made of the foreground star and the images of the 
aligned background star instead, which may be more practically referred to as
``spectroscopic differentiation of blending". Chwolson did not consider the 
lensing effects on the fluxes of the images and also concluded with a sentence 
``Whether the case of a fictitious double star considered here actually occurs,
I can not judge."

Then, it is very curious what Chwolson might have imagined for the flux of the
ring image of which he did not make any statements. If he presumed the same flux
as that of the unlensed background star as he did in his article for the (faint)
image, what must he have thought of the flux density around the ring? Distribute
two background star fluxes around the ring because two images turn into a ring
image? Distribute infinitely many point source stars around the ring? Which will result in an infinitely bright point star to an observer with a poor resolution
detector such as human eyes? Or, then, distribute finitely many background stars
around the ring because the stars have finite sizes? Then, what was the size of
the ring? What if the ring size is not an integer multiple of the size of the 
stars -- which in fact will be mostly the case apart from very rare coincidences?
We do not have access to Chwolson's article and can only guess from Schneider 
et al. that Chwolson must have recognized the axial symmetry but did not have 
enough details or interest to determine (or conjecture) the radius of the ring
image. And, Chwolson was not aware of the lensing effect on the image fluxes. 
Then, it is possible that Chwolson's remark on the ring image was mainly to
point out the deviated situation from the case of a spectroscopic double star of
which the latter was of his interest.

It seems to be quite reasonable to use the term {\it Einstein ring radius}
since Einstein was the first author to calculate the radius of the circular 
image. Einstein wrote in 1936\,\footnote{
``Lens-Like Action of a Star by the Deviation of the Light in the Gravitational
Field", Science, 84 (1936), p.506.}, ``An observer will perceive ... instead of
a point-like star A, a luminous circle of the angular radius $\beta$ around the
center of B, where $\beta = \sqrt{\alpha_0 R_0/D}$." Einstein calculated the 
ring radius in an approximation where the distance $D$ to the lensing star
({\it e.g.}, lensing by the Sun) is much smaller than the distance to the 
source star. ($\alpha_0$ is the deflection angle of a photon trajectory grazing
the surface of the foreground star and $R_0$ is the radius of the foreground 
star.) Einstein calculated the total magnification of the images (the second
equation in the article) and knew that the ring image is luminous. He wrote,
``This apparent amplification $q$ ... is a most curious effect, not so much for
its becoming infinite, ... ", and didn't seem to have concerned himself with 
the formal divergence. Physical objects are never a point and physical 
quantities are never exactly a delta function. In fact, Dirac delta function 
exists as a distribution which is defined within the context of integration. The
third equation in the Science article shows $q = 1/x\, (x\rightarrow 0)$ where
$x$ is the angular distance between the stars A and B. Integration of the 
apparent amplification over a small disk ${\cal D}$ centered at $x=0$
\opeqn
 \int_{\cal D} \Sigma(x_1, x_2)\,q\,dx_1 dx_2 
     = \int \Sigma(x_1, x_2)\,q\,xdx 2\pi
\cleqn
is well-defined and finite as far as the stellar surface flux density 
$\Sigma(x_1, x_2)$ is regular, which indeed is the case.

Then, how do we recognize that Chwolson was aware of and wrote in 1924 that the
image would be a ring centered at the foreground star when the foreground and
background stars are aligned (always meaning as seen from the observer with 
$G=0$)? Should we make the recognition by calling the circular image Chwolson
ring instead of Einstein ring? We consider a few aspects before casting an
intellectually reasonable vote.

\begin{itemize}
 \item
 Chwolson ring refers to a circular image. Einstein ring also refers to a 
circular image. (Precisely speaking, two half-circle images which should be
distinguishable by putting trace markers on the background star radiation 
pattern.) Chwolson mentioned it in 1924, and Einstein calculated it 12 years
later in 1936.

 \item
Chwolson ring is an image ring with unspecified radius and unspecified 
characteristics. Einstein ring is an image ring with a specified radius and
flux characteristics determined from the involved physical variables.

 \item
It seems most reasonable to call $\alpha_E$ the Einstein ring radius because
Einstein is the one who first recognized the importance of the radius and 
calculated it. Now if we choose to call a ring image a Chwolson ring, the
radius of the Chwolson ring may be most naturally called Chwolson ring radius.
If Chwolson had studied the characteristics of the image ring (such as the 
flux) establishing its physical nature and simply left out the radius in his
article to be supplemented by, say, Einstein 12 years later, it would be 
reasonable to call the image ring Chwolson ring and its radius Chwolson-Einstein
radius. But Chwolson did not. Considering the historical facts above, then,
it may be reasonable to call the image ring Chwolson-Einstein ring and the 
ring radius Einstein ring radius.  

 \item
For general lenses, Einstein ring and Einstein ring radius of a mass do not 
refer to the shape and radius of an image (or conjoined two images) but refer to
an imaginary circle and its radius that function as a scale disk where the 
functionality derives from the Einstein ring formula Einstein derived for the
first time (albeit for a case where the distance to the lensing star is small).
Since Chwolson ring is an image ring without physical functionality, it seems 
too big a leap, for example, to call the Einstein ring of the total mass of a 
binary lens Chwolson ring. 

 \item
Thus, we conclude that it is most reasonable to call Einstein ring and Einstein
ring radius Einstein ring and Einstein ring radius in their most general
contextual significances as is the current practice. On the other hand, it 
seems reasonable to call an observational circular image Chwolson-Einstein ring
and its ring radius Einstein ring radius. 
(``Circular images" are being found in cosmological 
lensings by galaxies -- extended mass distributions whose surface densities
are not necessarily circularly symmetric, and the ``circular images" are not
exactly circular even when the emission source is centered at the center of 
the caustic.) 

\end{itemize}

\subsection {The Linear Differential Behavior of the Lens Equation}

The lens equation in eq.\,(\ref{eqLens}) is written in terms of the variables
defined in the orbital plane $\theta=\pi/2$. In order for the lens equation 
to describe the  lensing corresponding to photon trajectories in an arbitrary
orbital plane, the lensing variables should be expressed in terms of  the two
dimensional variables defined in the observer's sky. If $x$, $s$, and $y$
are the two dimensional angular positions of an image, its source, and the lens,
the lens equation is given by
\opeqn
 x - s = \alpha_E^2 {x - y \over (x-y)^2} \ .
 \cleqn
In terms of complex coordinates,
\opeqn
 \omega = z -  {\alpha_E^2 \over \bar z - \bar x_1} \ ; \qquad ( \alpha_E = 1)
 \cleqn
 where $z$, $\omega$, and $x_1$ are the complexifications of 
 $x$, $s$, and $y$.  It is convenient and customary to normalize the lens
 equation so that $\alpha_E =1$ (as is indicated in the parenthesis).
 
The lens equation is an explicit  function from a two-dimensional  image
variable to a two-dimensional source variable. In other words, the lens 
equation is a mapping from a complex plane to itself. The Jacobian matrix
of the lens equation describes its linear differential behavior, and when the 
Jacobian determinant vanishes, the dimension of the  vector space of the mapped 
decreases. In other words, an infinitesimal two-dimensional image is mapped to 
an infinitesimal source of one-dimension. (The trace of the Jacobian matrix is 
non-zero, and when one eigenvector vanishes, the  other is non-zero -- in fact, 
2.) The set of points where the Jacobian determinant $J$ vanishes is called 
the critical curve. 
  \opeqn
     J = 1 - |\kappa|^2 = 1 - {1\over |z-x_1|^4}  \ ; \qquad
         \kappa \equiv {\partial \bar \omega\over \partial z}
  \cleqn
The critical curve ($J=0$) is a circle of radius 1 ($1=\alpha_E$ or $R_E$,
usually, depending on how to normalize the lens equation) centered at $ x_1$.
In other words, the critical curve coincides with the circular image of the
Einstein ring. The critical curve is mapped to $\omega = x_1$, hence the lens 
position is the position of the point caustic. Inversely, a point source at the
lens position produces the circular image on the critical curve. 
  
It is useful to define linear Einstein ring with radius 
$R_E \equiv D_1 \alpha_E$ ( mentioned above).
The photon trajectories that form the circular image passes through the (linear)
Einstein ring at their closest approaches to the mass $M$; in other words, their 
periastron distances is $r_0 = R_E$.
  
  The Jacobian matrix
  \opeqn
    {\cal J} = \pmatrix{ 1  \  \bar\kappa \cr
                                     \kappa  \  1 }
\cleqn
has eigenvalues $\lambda_\pm = 1 \pm |\kappa|$, and the eigenvectors are
\opeqn
  e_+ = \pmatrix{E_+ \cr \bar E_+} \  ; \quad
  e_- = \pmatrix {E_- \cr \bar E_-} 
  \cleqn
  where
  \opeqn
   \kappa = |\kappa| e^{-i2\theta} \ ; \qquad  E_+ \equiv e^{i\theta} \ ;  \qquad E_- \equiv  i E_+ \ .
 \cleqn
On the critical curve, $|\kappa|=1$ and  $\lambda_-=0$, and the critical 
eigenvector $\pm E_-$ is tangent to the critical curve at every critical point. 
(Eigenvectors are not assigned the senses. Thus, $\pm E_-$.)
In other words, every critical point is a precusp, and the point caustic is 
a degenerate cusp.

\subsubsection{Breaking the Degeneracy with a Small Constant Shear}
\label{secBreak}

We break the degeneracy of the point caustic of a point mass by introducing a 
small constant shear to understand the degeneracy as a limit. Recall how to 
produce a constant electric field using a dipole where two large opposite 
charges are separated by a large distance. We can introduce a large mass at a 
large distance such that $mass/distance^2$ is constant, to a similar effect.
If the point mass is at $z=x_1$ where $x_1=0$ is real, and the large mass is
on the negative real axis, the lens equation is given by 
\opeqn
 \omega = z + \epsilon \bar z - {1\over \bar z} \; \qquad 
          \epsilon \geq 0 \ .
\label{eqCR}
\cleqn
For our purpose, we need small $\epsilon$. The lens equation (\ref{eqCR})
known as Chang-Refsdal model\,\footnote{
K. Chang and S. Refsdal, 1979, Nature, 282, 561.} has a bifurcation of the
critical curve (hence also caustic curve)\footnote{
{\it Ibid.}, A \& A, 132, 168.} at $\epsilon = 1$, hence we assume 
$\epsilon < 1$ so as to take a smooth limit of $\epsilon \rightarrow 0$.
\opeqn
 \kappa = \epsilon + {1\over z^2} \equiv |\kappa|\,e^{-2i\theta}
\cleqn
and the critical curve is where $|\kappa| =1 $. If $z$ is real, then $\kappa$ is
real and positive, hence $\kappa=1$ where the critical curve intersects the real
axis (or the ``dipole axis"). The critical points on the ``dipole axis" are
\opeqn
 z = \pm (1-\epsilon)^{-1/2} \ ,
\cleqn
and they exist for $\epsilon < 1$. The critical eigenvector at the critical 
points are $\pm E_- = \pm i e^{i\theta} = \pm i$ since 
$\kappa = 1 \ (\theta = 0,\ \pi)$. The tangent to the critical curve at the 
critical points,  
\opeqn
 {dz\over d\theta} = \pm i\,(1-\epsilon)^{-3/2} \ \propto \ \pm E_- \ ,
\cleqn
are parallel to the critical eigenvector $\pm E_-$, and so the critical points
are precusps. The corresponding cusps are at
\opeqn
 \omega = 2\epsilon z = \pm 2\epsilon (1-\epsilon)^{-1/2} \ .
\label{eqCuspR}
\cleqn

The lens equation has only one singularity (pole) at $z=0$ and its topological
charge is 1. There are two limit points where $\kappa =0$, namely, at
$z = \pm i \epsilon^{-1/2}$. Thus, the critical curve is made either of one 
loop enclosing the pole ($\epsilon < 1)$ or of two loops each enclosing a limit
point ($\epsilon > 1)$. One-loop critical curve has topological charge 1 and 
produces a 4-cusp caustic. Each loop of the two-loop critical curve has 
topological charge $1/2$ and produces a triangular caustic. The bifurcation from
one quadroid to two trioids occurs at $\infty$. Here we are interested in the 
quadroid because it is the quadroid that contracts to a point in the limit
of a point mass lens $\epsilon \rightarrow 0$. The quadroid has two cusps on
the real axis given in eq.\,(\ref{eqCuspR}) and it is easy to guess correctly
from the symmetry that the other two are on the imaginary axis, reflection 
symmetric with respect to the ``dipole axis". The quadroid is bisected by the
real axis.

Now consider $\omega$ on the real line. There are two real solutions for $z$.
\opeqn
 0 = (1+\epsilon) z^2 - \omega z - 1
\cleqn
For $|\omega| > 2\epsilon(1-\epsilon)^{-1/2}$, one solution is negative
($J\leq 0$) and the other is positive ($J\geq 0$). 
For $|\omega| > 2\epsilon(1-\epsilon)^{-1/2}$, both are negative ($J < 0$).
The lens equation (\ref{eqCR}) can be embedded into a fourth order polynomial
equation for $z$, hence there can be two more solutions. Substitute 
$z = r e^{i\phi}$ and find that the solutions are on a circle centered at the
position of the point mass lens element.
\opeqn
 r^2 = {1\over 1-\epsilon} \ ; \qquad \omega = 2\epsilon r \cos\phi
\cleqn
The solutions exist for $|\omega| \leq 2\epsilon(1-\epsilon)^{-1/2}$, which 
defines the real line segment contained by the two cusps; Inside the quadroid,
there are two extra images for each $\omega$ and they are on the circle of 
radius $(1-\epsilon)^{-1/2}$. They are positive images ($J\geq 0$). The two
extra images satisfy the following quadratic equation, which can be found by
dividing the fourth order polynomial by the quadratic equation for the real
solutions.
\opeqn
 0 = z^2 - \omega \epsilon^{-1} z + (1-\epsilon)^{-1}
\cleqn
where $\omega$ is real and inside the quadroid caustic. 

\begin{figure}
\plotone{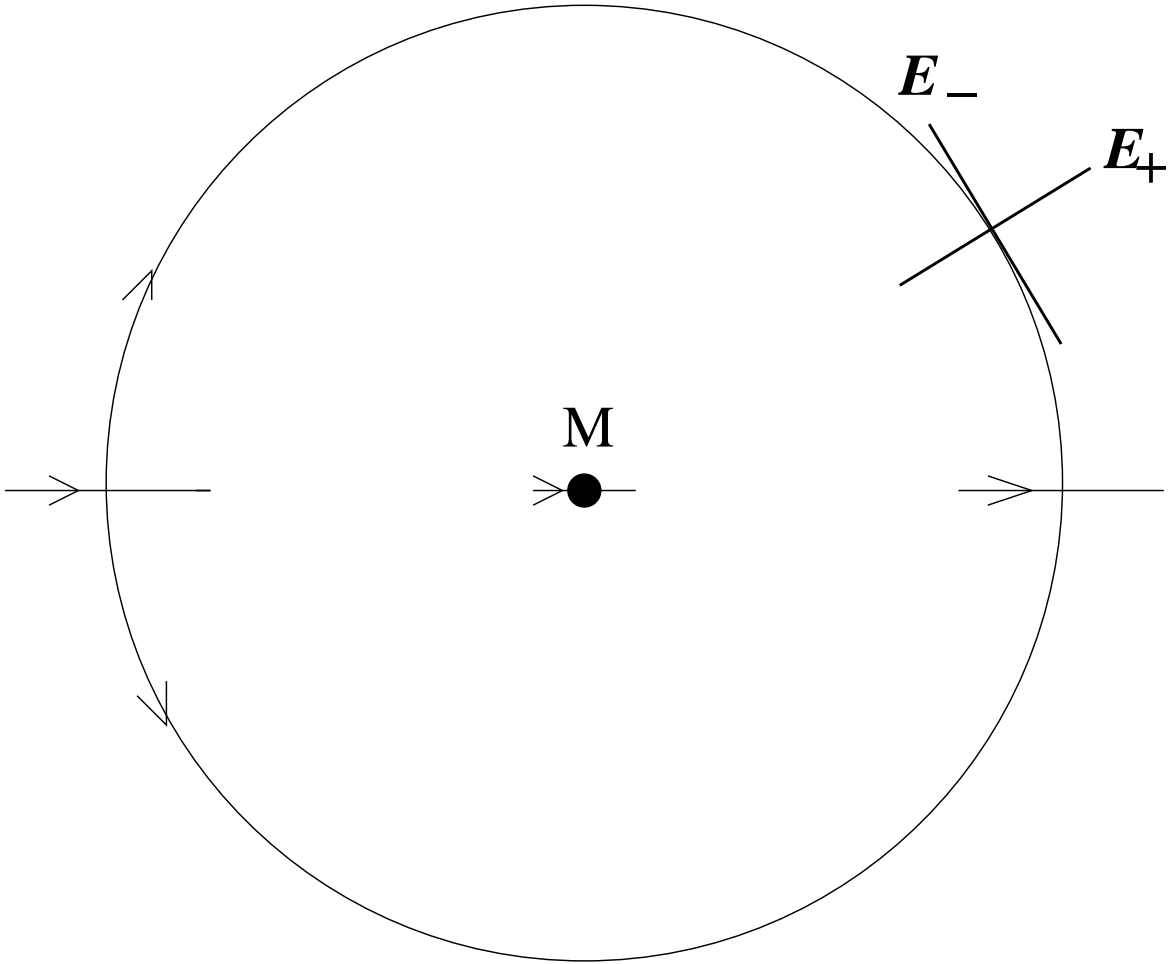}
\caption{Einstein ring as the location of the ring image and critical curve.
$\pm E_+$ is the positive eigenvector (for $\lambda_+$) and radial, and 
$\pm E_-$ is the negative eigenvector (for $\lambda_-$) and tangential.
As the source crosses the point caustic along the real line, two images on
the real line reach two critical points, then two images instantaneously 
trace the Einstein ring at the crossing as the two arrows on the ring 
indicate, and two images move on along the real line. The last: one 
toward the lens position and the other toward $\infty$.} 
\label{how_fig}
\end{figure}

If we consider $\omega$ moving on the real line in the positive direction and
crossing a cusp $\omega = - 2\epsilon(1-\epsilon)^{-1/2}$, the positive image on
the real line crosses the critical point $z = - (1-\epsilon)^{-1/2}$ into the
area enclosed by the critical curve and turns into a negative image. The other
image on the real line moves in the positive direction maintaining its 
(negative) parity until $\omega$ crosses the other cusp
$\omega = 2\epsilon(1-\epsilon)^{-1/2}$. In the  mean time, the two extra images
trace the circle: one image, the half-circle in the upper half plane, and the
other image, the half-circle in the lower half plane. Figure\,\ref{how_fig} is
a depiction of the Einstein ring as the critical curve of the single point mass
lens $M$. Let's consider the ring, for a moment, as the circle of the two extra
images (and imagine the quadroid centered at $M$): the arrow near the point mass
lens depicts the motion of the source $\omega$; two arrows on the real line 
depict the images moving on the real line; two arrows on the ring depict the two
extra images that start at one precusp and end at the other precusp.

Now take the limit $\epsilon \rightarrow 0$, and one can visualize two images
tracing the Einstein ring instantaneously at the crossing of the point caustic.

We find the above a comfortable (usually involving continuity or traceability) 
way to think of the transition from two point images to two half-circle images.
The criticality of a lens equation in general is related to formation or 
disappearance of the two extra images (or higher even number of images). In the
case of the single point mass lens, the two extra images trace a ring 
instantaneously forming a ring image due to the degeneracy, and the dimensional
change of the linear differential vector spaces of the lens equation is 
manifested in a global manner.

 \section { Caustics of Lensing by Slowly Moving Matter }
  
The exquisite symmetry of the Schwarzschild  lens (spherically symmetric matter)
that led to the caustic of degenerate cusp is readily broken in a more general 
lens, and the caustic is generally a one-dimensional curve as is the critical 
curve. The lens equation obtained with the  small angle approximation is a 
mapping of the complex plane (extension of a small neighborhood of the 
observer's sky where the small angle approximation is valid) to itself. The 
criticality condition $J=0$ imposes one constraint reducing the dimension of 
the variable space by one. Thus, the critical curve is a one dimensional curve 
which may be a disjoint sum of many loops, and they are usually smooth. Also, 
the lens equation is usually smooth almost everywhere except at the poles due 
to the point mass lens elements ($z=x_1$ in the case of the single point mass
lens discussed above). At the poles, $J =-\infty$, hence it is reasonable to 
assume that the lens equation is smooth on the critical curve for gravitational 
lenses in general.
 
A smooth curve is mapped to a smooth curve by a smooth mapping except at the 
stationary points of the mapping. Recall that one of the two eigenvectors 
vanishes on the critical curve. If we consider a small deviation $dz$ from 
a critical point,
 \opeqn
    dz = dz_+ E_+ + dz_- E_- \qquad \rightarrow \qquad
   d\omega = d\omega_+ E_+  \ ,
\cleqn
hence $\omega$ changes only in the direction of $E_+$. Now if $dz_+=0$, then
$d\omega=0$. So, if we imagine tracing the critical curve (or integrating the 
tangent to the critical curve) and mapping to draw the caustic curve, 
the procession of the caustic curve stops momentarily where
$dz_+=0$ and turns around. In other words, the caustic curve develops a cusp
where the tangent to the critical curve is parallel to the critical eigenvector 
$\pm E_-$.  The critical point where the tangent has $dz_+=0$ is called precusp.
Thus, the caustic curve of a lens equation is  a piecewise smooth curve
punctuated by cusps which may be a disjoint sum of cuspy loops that may
intersect themselves. 

As the point caustic at the lens position of the single lens is defined
within the context of the lens equation, the caustic curve of a general 
gravitational lens is defined only within the context of the lens equation.

\section { Comments on Two Papers}

We discuss a couple of papers found in the Astrophysical Journal that may play
an unfortunate role of perpetuating misunderstandings.

\subsection {``Superluminal Caustics" in 2002}

 The paper entitled ``Superluminal Caustics of Close, Rapidly-Rotating
Binary Microlenses", Zheng Zheng and Andrew Gould, ApJ 541 (2002), 728,  
considers the caustic curve of the binary lens equation as an object of the 
size given by the multiplication of its angular size and distance $D_1$ and 
of an (unspecified) inertia whose velocity has to be compared with the speed 
of light and be concerned of for its tachyonic nature. As we discussed in this 
article, the caustic is  defined only within the context of the lens equation, 
and especially, it is not an object of an inertia. There is nothing wrong with 
thinking of the caustic curve as a large linear object by projecting it to the 
normal plane passing through the source or passing through the lens, or at any 
distance along the line of sight if that serves to understand or apply the lens
equation for some purpose as far as the contextual existence and 
characteristics of the caustics as defined within the lens equation are valid. 
Certainly, inertia is not a quality of the caustics defined in the lens 
equation; There does not exist a tachyonic caustic.

If we stipulate for a moment for the purpose of clearing up another conceptual 
mistake that the caustics be endowed with inertias, what the caustics can newly 
acquire is the tachyonic nature but not a superluminal phenomenon known in 
astrophysics as we discussed in astro-ph/0002414. Zheng and Gould speculate to 
detect their  ``superluminal caustics" using very large telescopes such as 
of 100m aperture. We repeat that tachyonic caustics do not exist; 
The nomenclature of  ``superluminal caustic" is a misidentification even
within the context of their (mistaken) idea about the motion of the caustics. 
There does not exist a physically perceivable notion of  
``superluminal caustic" nor a tachyonic caustic, and it would be an unnecessary 
waste of resources to consider detecting ``superluminal caustics" 
with say LHT (Larger than Huge Telescope).

Other serious problems of the paper by Zheng and Gould can be found elsewhere.

How is such a paper published in the Astrophysical Journal in 2002?  Perhaps, 
in the same way other worthy papers are smothered in the referee system.

\subsection {``Fermat's Principle" in General Relativity in 1990}

 The paper entitled ``Fermat Principle in Arbitrary Gravitational Fields", 
Israel Kovner, ApJ 351 (1990), 114, is cited in the monograph on lensing 
titled ``Gravitational Lenses" by Schneider et al. for Fermat's principle.  
Thus, it is likely that the Kovner's article is influential. Among others,
we discuss why ``the least proper time principle", a generic variational 
principle in general relativity, should not be referred to as Fermat's 
principle. 

\subsubsection 
{Focal Points of a Gravitational Spherical Geometry are not Caustics}

Kovner seems to misidentify the source and observer as the caustics (and
``past caustics") as he states in the second paragraph of section IV, p.\,118,
``The caustics are mergings of $\geq 2$ extremals of the emission time, ...
the {\it past}-caustics of the past light cone emanating from the 
observer ... ."  We discussed above hat the source and observer are the two 
focal points of a particular set of null geodesics, namely the null geodesics 
that connect the source and the observer. An observer can see a photon from 
the source only if the photon arrives at the detector (or the eye) of the 
observer, and it is imperative that there exist some null geodesics that
connect the source and the observer if the observer will be able to image the 
source photonically.  In plane geometry, there is only one (null) geodesic
between the two points defined by the source and the observer. In spherical
geometry, the geodesics cross, resulting in multiple geodesics connecting
two points. There is where the notion of the focal points comes in. They are
the focal points of the geodesics of the spherical geometry. Distinct null 
geodesics produce distinct images, the spherical geometry is responsible for
multiple images, and the source and observer are the two foci of the null
geodesics corresponding to the multiple images. 

The misidentification of the source and observer as caustics may derive from 
the definition of caustics in electromagnetic lensing. See, for example, ``The 
Classical Theory of Fields", L.\,D.\,Landau and E.\,M.\,Lifshitz, 1975, p.133.

\subsubsection{Least Action Principle in Non-Relativistic Mechanics}

``Fermat's principle" in general relativity is a four-dimensional version of the
least (abbreviated) action principle\footnote{
The least action principle seems to referred to as Maupertuis' principle in 
certain literature even though the formulation was due to Euler and Lagrange.
Goldstein (chap.\,7) writes, ``However, the original statement of the principle 
by (Pierre de) Maupertuis (1747) was vaguely theological and could hardly pass
muster today. The objective statement of the principle we owe to Euler and 
Lagrange." We gather that calling the principle of least action Maupertuis'
principle may amount to calling Fermat's principle Cureau's principle.}
of the non-relativistic mechanics where the lagrangian has only the kinetic
terms.

The principle of least action in non-relativistic mechanics states that the 
``abbreviated" action 
\opeqn
 S_0 = \int dt\, P\dot Q
\cleqn
of a system for which the Hamiltonian is conserved is the extremum along the 
equation of the motion. $Q$ is the coordinate variable of the particle and
$P$ is its conjugate. Recall that the equations of motion are obtained as the
extremum of the action
\opeqn
 S = \int dt \, L = \int P dQ - H dt \ .
 \label{eqS}
\cleqn
In a system with conserved Hamiltonian, one can consider only the paths that
satisfy the conservation of the Hamiltonian by constraining by the equation of
the motion and allowing a variation of the final time $t$ (or final time and
initial time: See Goldstein or Landau and Lifshitz\footnote{
{\it Mechanics}, L.\,D.\,Landau and E.\,M.\,Lifshitz, Pergamon press, 1969.}),
and the (abbreviated) action $S_0$ is the action that has an extremum under the
variation. If we denote the variation $\Delta$ variation after Goldstein
(chap.\,7),
\opeqn
 \Delta S = \left(L - {\partial L\over \partial \dot Q} \dot Q \right)
             \Delta t = - H \Delta t \ .
\cleqn
From the second equality of eq.\,(\ref{eqS}) and conservation of $H$, 
\opeqn
  \Delta S = \Delta S_0 - H \Delta t \ ,
\cleqn
hence $\Delta S_0 = 0$. Now if there is no external force and the kinetic 
energy is conserved, $S_0 = 2H \int dt$ and the particle follows a path such
that the time it takes is the extremum (usually minimum, hence least time 
principle). 
\opeqn
 0 = \Delta S_0 = 2 H \Delta t
\cleqn
It recalls Fermat's principle in geometric optics which is also called 
{\it the principle of least time}. 

In order to see the structural parallel (as a functional analysis) with the
four-dimensional case of the general relativity, note that Hamiltonian and
time variable are a pair of canonical conjugates.

\subsubsection {Least Action Principle in General Relativity}

The Hamiltonian of the system of a freely falling particle described in
eq.\,(\ref{eqAction}) is nothing but the (half the) momentum square of the
particle\footnote{
The scalar product of the four momenta $P^2 =M^2$ where $M$ is the mass of the 
particle, and Kovner refers to the set of $\{P_\mu\}$ that satisfies the 
condition the mass shell as is customary in particle and nuclear physics. Off
the mass shell states are known as virtual particles. Feynman diagrams are a
web of interactions of virtual and real particles that pictorially describe
combinatoric calculations of Feynman path integrals.}
and is conserved along the equation of motion. The Hamiltonian ${\cal H}$ and
the path parameter $p$ are canonical conjugates. There is no external force for
a particle freely falling in a curved space time as defined by the metric, 
hence this is an exact parallel with the non-relativistic case discussed above
where the kinetic energy is conserved. The resulting variational principle is a 
principle of extremum path parameter $p$, or equivalently a principle of 
extremum proper time $\tau$. Recall the action in eq.\,(\ref{eqAction})
and define $S_0$,
\opeqn
 S = \int {\cal L} dp \ ; \qquad  S_0 = \int P_\mu \dot x^\mu dp \ ,
\cleqn
and the $\Delta$ variation of the action $S$ is given by
\opeqn
 \Delta S = - {\cal H} \Delta p \ .
\cleqn
Since the Hamiltonian is conserved, 
\opeqn
  \Delta S = \Delta S_0 - {\cal H} \Delta p \ ,
\cleqn
hence $\Delta S_0 =0$.  From $S_0 = \int 2 {\cal H} dp$, \ 
$0= \Delta S_0 = 2 {\cal H} \Delta p$. Expressing in terms of the $\Delta$
variation of the proper time,
\opeqn
 0= \Delta S_0 = 2 {\cal H} \Delta p = - {d\tau^2\over dp^2}\Delta p
  = - {d\tau\over dp} \Delta\tau 
  = - (-g_{mu\nu}\dot x^\mu \dot x^\nu)^{1/2} \Delta \tau \ .
\cleqn
The space-time path taken by the particle is such that the proper time elapsed
is in extremum.
\opeqn
 0 = \Delta \int d\tau = \Delta (\tau - \tau_-) = \Delta \tau
\label{eqElapsedTau}
\cleqn
If we consider photons leaving a light source at $\tau = \tau_-$, the light 
paths are determined such that the arrival time $\tau=\tau_+$ ( for a given
$\tau_-$) or the travel time $\delta \tau = \tau_+ - \tau_-$ is an extremum.

One potential pitfall in eq.\,(\ref{eqElapsedTau}) is the case where 
$d\tau^2=0$ because, then, $0=\Delta S_0$ does not necessarily imply
$0 = \Delta \tau$. We may take the limit $d\tau^2\rightarrow 0$ as the case 
for photons and assume that $0 = \Delta \tau$ is valid. Then, we need to 
confirm that the new action $\int d\tau$ does generate the equations of motion
as its extremum.
\opeqn
 S_\tau = \int d\tau = \int {d\tau\over dp} dp
        = \int (-g_{\mu\nu}\dot x^\mu \dot x^\nu)^{1/2} dp
\label{eqActionTau}
\cleqn
Under the variation of the path $x^\mu(p)$, the variation of the action is
(Weinberg, p.76)
\opeqn
  \delta S_\tau = - \int \left({d^2 x^\alpha\over d\tau^2}
     + \Gamma^\alpha_{\mu\nu}{dx^\mu\over d\tau}{dx^\nu\over d\tau}\right)
    g_{\alpha\beta}\, \delta x^\beta\, d\tau \ .
\cleqn
If a particle follows the equations of free fall, the variation of the action
vanishes. In other words, the physical space-time path a particle follows is
such that the proper time elapsed is an extremum. Now, there is no ambiguity
related to $d\tau^2=0$; Photons indeed follow paths for which the travel times
(or arrival times for given departure times) are extrema.

\begin{enumerate}
 \item
The last paragraph in p.100 of the monograph ``Gravitational Lenses" by
Schneider et al. states, `` ... does not refer to the `time' a light ray needs
to travel from the source to the observer ... but a stationary property of the
(invariant) {\it time of arrival} at the observer ... ." Since the arrival 
time $\tau = \tau_+$ and the travel time $\delta \tau = \tau_+ - \tau_-$ are
effectively the same variables once given the time of the departure
$\tau = \tau_-$, it is a self-contradiction. If the arrival time can be defined,
so can the travel time. If the arrival time along a geodesic is an extremum, so
is the travel time along the same geodesics. Then, why do Schneider, Ehlers,
and Falco {\bf denounce} the notion of travel time while accepting arrival time?
It is unclear within the section on ``Fermat's principle" in the book. The 
source of the error may be $\delta \lambda_{obs}=0$ in the Kovner's article 
where Kovner does not elaborate the new action $S_0$. ($\lambda$ is the path 
parameter in Kovner's.) It is apparent that the error has propagated unchecked.

 \item
What is conspicuous is that eq.\,(\ref{eqActionTau}) is just another commonly 
used path integral and variation in general relativity to derive the equations
of motion from a variational principle (Weinberg, p.76). The freely falling 
particle can be massive or massless. For the latter, $d\tau^2 = 0$, and for 
the former, $d\tau^2 \neq 0$ which we may normalize such that $d\tau^2=1$.
Now, should we pluck out, of the continuum (at least classically) of mass 
spectrum, the massless subset given by $d\tau^2=0$ and call it Fermat's 
principle?  Most certainly not. 

 \item
The evolution of the fundamental physics has been in the direction of 
unifications. Not only the light but a car is a wave according to quantum
mechanics except that the latter has much shorter matter (or de Broglie) 
wave length. In the limit of zero wavelength, the classical equations are
recovered either for massless particles (photons, phonons, etc) or massive
particles. Maxwell equations for light lead to the eikonal equation (or
geometric optics) in the limit of zero wavelength that is suitable to 
describe the propagation of the wave front of the light bundle in a medium
whose electromagnetic properties vary slowly -- usually, wrapped up as a 
slowly varying refraction index. The eikonal equation can be derived from an
action and its variation. It corresponds to a case where the kinetic energy
is conserved. Thus, Fermat's principle, or the principle of least time.
(The historical account of Fermat's principle can be found in a separate 
article.)

In general relativity, unlike in Newtonian gravity, the gravitation is not given
as an external force but is integrated into the general covariance. It is
exactly in the same manner as the electromagnetic forces are integrated into
$U(1)$ gauge covariance, for example, in quantum electrodynamics. In this
framework of ``geometrization", the forces are integrated into dynamic 
variables. As a result, the lagrangian ${\cal L}$ in eq.\,(\ref{eqAction})
describes a free particle, has only the kinetic terms, which is conserved,
and allows simple variations.

 \item
Geometric optics (or eikonal equation) is exactly valid in the limit of zero
wavelengths and can be described by the geodesic equation when under gravity.
Huygens principle is about diffraction phenomenon where the wavelengths of the
light with respect to the aperture matter: ``Huygens wavelets" propagate out
as small spherical waves from every point of the aperture (with a subtle
understanding of the nature of the approximation that the point is 
sufficiently smaller than the aperture and is sufficiently larger than
the wavelength). It is unclear how the ``Huygens wavelets" are related to 
light cones\footnote{Light cones of a space-time point $x_0$ is the set 
$\{x^\mu\}$ that satisfies $(x-x_0)^2=0$. In a sufficiently small neighborhood 
of $x_0$, the set is made of two back-to-back ``cones" connected at $x_0$. 
For a massive particle, the future and past hyperboles are separated by the 
mass gap.} and ``Fermat's principle" as Kovner states in p.116, 
``Another way to regard the Fermat principle for light
is provided by the Huygens principle as illustrated in Figure 2." Does Kovner
imply that diffractions are also described by geometric optics? 

\end{enumerate}

\end{document}